\documentclass[aps,showpacs,nofootinbib]{revtex4}

\usepackage{graphicx}
\usepackage{graphics}
\usepackage{amssymb}
\usepackage{amsmath}

\newcommand{\half}{\textstyle{\frac{1}{2}}}

\newcommand{\be}{\begin{equation}}
\newcommand{\ee}{\end{equation}}
\newcommand{\bea}{\begin{eqnarray}}
\newcommand{\eea}{\end{eqnarray}}
\newcommand{\beal}{\begin{align}}
\newcommand{\eal}{\end{align}}
\newcommand{\bespl}{\begin{split}}
\newcommand{\espl}{\end{split}}
\newcommand{\nn}{\nonumber}
\newcommand{\nslash}{\kern 0.2 em n\kern -0.50em /}
\newcommand{\kslash}{\kern 0.2 em k\kern -0.45em /}
\newcommand{\pslash}{\kern 0.2 em p\kern -0.50em /}
\newcommand{\Sslash}{\kern 0.2 em S\kern -0.50em /}
\newcommand{\Pslash}{\kern 0.2 em P\kern -0.50em /}
\newcommand{\Rslash}{\kern 0.2 em R\kern -0.50em /}

\begin{document}
\title{
Monte-Carlo simulation of events with Drell-Yan lepton pairs from 
antiproton-proton collisions}

\author{A.~Bianconi}
\email{andrea.bianconi@bs.infn.it}
\affiliation{Dipartimento di Chimica e Fisica per l'Ingegneria e per i Materiali, 
Universit\`a di Brescia, I-25123 Brescia, Italy, and\\
Istituto Nazionale di Fisica Nucleare, Sezione di Pavia, I-27100 Pavia, Italy}

\author{Marco Radici}
\email{radici@pv.infn.it}
\affiliation{Dipartimento di Fisica Nucleare e Teorica, Universit\`{a} di Pavia, 
and\\
Istituto Nazionale di Fisica Nucleare, Sezione di Pavia, I-27100 Pavia, Italy}

\begin{abstract}
The complete knowledge of the nucleon spin structure at leading twist requires also
addressing the transverse spin distribution of quarks, or transversity, which is
yet unexplored because of its chiral-odd nature. Transversity can be best 
extracted from single-spin asymmetries in fully polarized Drell-Yan processes with 
antiprotons, where valence contributions are involved anyway. Alternatively, in 
single-polarized Drell-Yan the transversity happens convoluted with another 
chiral-odd function, which is likely to be responsible for the well known (and yet 
unexplained) violation of the Lam-Tung sum rule in the corresponding unpolarized 
cross section. We present Monte-Carlo simulations for the unpolarized and 
single-polarized Drell-Yan $\bar{p} p^{(\uparrow )} \rightarrow \mu^+ \mu^- X$ at
different center-of-mass energies in both configurations where the antiproton beam hits
a fixed proton target or it collides on another proton beam. The goal is to estimate 
the minimum number of events needed to extract the above chiral-odd distributions from 
future measurements at the HESR ring at GSI. It is important to study the 
feasibility of such experiments at HESR in order to demonstrate that interesting 
spin physics can be explored already using unpolarized antiprotons.
\end{abstract}


\maketitle

\section{Introduction}
\label{sec:intro}

When studying the spin structure of hadrons, the best tool is represented by the
spin asymmetry, defined by the ratio between the difference and the 
sum of differential cross sections obtained by flipping the spin of one of the
involved polarized hadrons. For hadronic beams, spin asymmetries are known since a 
long time; historically, the first one was obtained at FERMILAB, where an 
anomalous large transverse polarization of the $\Lambda$ produced in 
proton-nucleon annihilations was measured~\cite{Bunce:1976yb}, surviving even at 
large values of the $\Lambda$ transverse momentum. More recently, similar anomalously
large asymmetries have been observed by the FNAL E704~\cite{Adams:1991cs} and 
STAR~\cite{Adams:2003fx} collaborations in inclusive pion production from collisions 
of transversely polarized protons. Such asymmetries require a nonvanishing 
imaginary part in the off-diagonal block of the fragmentation matrix of quarks into 
the detected hadron, which is forbidden in perturbative QCD at leading 
twist~\cite{Kane:1978nd}. Soon after the first FERMILAB observation, a pioneering 
work appeared~\cite{Ralston:1979ys} about the possibility of having leading-twist 
asymmetries in fully polarized Drell-Yan processes, based on a new partonic 
function describing the quark transverse spin distribution in a transversely 
polarized hadron: the transversity. However, this work was basically ignored 
(and the observed asymmetries left unexplained) upon the prejudice that 
transverse-spin effects have to be suppressed. Only almost a decade later, it was 
realized that the transversity is connected to a helicity flipping mechanism but 
it is still a leading-twist distribution, necessary to complete the knowledge of 
the nucleon spin structure~\cite{Artru:1990zv,Jaffe:1991kp} (for a review, see 
also Refs.~\cite{Jaffe:1996zw,Barone:2003fy}).

As such, the transversity is not diagonal in the parton helicity basis; since at 
leading twist helicity and chirality coincide, it is usually mentioned as a 
chiral-odd function. But in the transverse spin basis it is diagonal and can be 
given a probabilistic interpretation. To be precise, in a transversely polarized 
hadron it describes the probability difference for having a transversely polarized 
parton with spin parallel or antiparallel to the one of the parent hadron. The 
chiral-odd feature brings in negative and positive consequences. The transversity 
needs a chiral-odd
partner in order to be measured in a (chiral-even) cross section. At leading twist, 
the partner can be the transversity itself (for the corresponding antiquark in the 
aforementioned polarized Drell-Yan) or another chiral-odd unknown fragmentation 
function in a semi-inclusive process. In particular, the transversity is not
accessible in inclusive Deep-Inelastic Scattering (DIS)~\cite{Jaffe:1993xb}. 
Therefore, in the Quark Parton Model (QPM) description of DIS, it does not have a 
counterpart at the level of structure function. Even if it depends on spin, it is 
not related to some partonic fraction of the hadron spin; the related twist-2 
tensor operator $\sigma^{\mu \nu} \gamma_5$ is not part of the hadron full angular 
momentum tensor, and its first moment, the tensor charge, has a 
nonvanishing anomalous magnetic moment: contrary to the nonsinglet axial charge, 
from a starting input scale the tensor charge unavoidably evolves to 
zero~\cite{Jaffe:1996zw}.

If the hadron target has spin $\half$, there is no transversity for gluons because
of the mismatch in the change of helicity units. Hence, during evolution the quark
transversity decouples from any radiative gluon. Moreover, it changes sign under 
charge conjugation, which implies that again during evolution it does not receive
any contribution from charge-even structures like the $q\bar{q}$ pairs from the
Dirac sea. In summary, the very same features that make the transversity quite an
elusive object, are responsible for its very peculiar behaviour. Its evolution is
similar to the one of a nonsinglet structure function; it describes the spin
distribution of a valence quark while "switching off the contribution of gluons and
of the vacuum"~\cite{Jaffe:1996zw}. 

The growing interest in the transversity reflected in a rich experimental program, 
that led to some first recent measurements with hadronic~\cite{Bravar:2000ti} and 
leptonic 
beams~\cite{Airapetian:2000tv,Airapetian:2001eg,Airapetian:2002mf,Airapetian:2004tw}. 
However, the field is confronted with rapid developments that should bring new
upgraded and precise results in the near future~\cite{Elschenbroich:2004ba}. 
Transverse spin is intimately connected with an explicit dependence of the parton 
distribution on its intrinsic transverse momentum. Including this explicit 
dependence, the list of parton distribution (and fragmentation) functions is very 
rich already at leading twist, containing also other chiral-odd 
objects~\cite{Mulders:1996dh}. While on the theoretical side the discussion is 
focussed on extending the universality proof to such 
functions~\cite{Boer:2003cm,Collins:2002kn,Metz:2002iz,Bomhof:2004aw,Collins:2004}, 
new azimuthal asymmetries are being deviced and measured both on the 
phenomenological~\cite{Yuan:2003gu,Gamberg:2003pz,Bacchetta:2004zf}
and experimental~\cite{Avakian:2003pk} side, that can shed light on new underlying 
mechanisms, ultimately related to the orbital motion of the quarks in the parent 
hadron. 

In principle, the aforementioned polarized Drell-Yan is the most convenient process
to measure the transversity at leading twist, because it does not involve other
unknown functions. However, a Next-to-Leading (NLO) simulation has shown that the 
QCD evolution and the Soffer inequality (if forced to be fulfilled at any scale 
during evolution) make the resulting spin asymmetry very 
small~\cite{Martin:1998rz,Barone:1997mj}. Moreover, the latter involves the 
transverse spin distribution of an antiquark in a transversely polarized proton, 
which is not expected to be large. This difficulty can be overcome at the foreseen 
HESR ring at GSI, where (polarized) antiprotons will be 
produced~\cite{pax,assia,Efremov:2004qs,Anselmino:2004ki,Radici:2004ij}: in 
the process $\bar{p}^\uparrow p^\uparrow \rightarrow l^+ l^- X$, only valence 
(anti)quark distributions are involved without any further suppression. 

But interesting information on the partonic spin structure of hadrons can also 
be inferred by using unpolarized antiproton beams only. In fact, in 
single-polarized Drell-Yan like $\bar{p} p^\uparrow \rightarrow l^+ l^- X$ the 
polarized part of the cross section contains at leading twist three terms of which
two produce interesting azimuthal spin 
asymmetries~\cite{Boer:1999mm}. The first one involves the convolution of the 
standard unpolarized distribution $f_1$ with the Sivers function 
$f_{1T}^\perp$~\cite{Sivers:1991fh}, which describes how the distribution of 
unpolarized quarks is affected by the transverse polarization of the parent proton. 
The very same $f_{1T}^\perp$ appears at leading twist in semi-inclusive DIS 
processes like $lp^\uparrow \rightarrow l' \pi X$~\cite{Efremov:2003tf} or 
$pp^\uparrow \rightarrow \pi X$~\cite{Anselmino:2004??}, and it is responsible of 
an azimuthal asymmetric distribution of the detected pions depending on the 
direction of the target polarization, the socalled Sivers effect. A measurement of 
a nonvanishing asymmetry would be a direct evidence of the orbital angular 
momentum of quarks~\cite{Elschenbroich:2004ba}. 

The second term in the single-polarized Drell-Yan cross section has a different
azimuthal dependence and it involves the transversity $h_1$ convoluted with the 
chiral-odd distribution $h_1^\perp$~\cite{Boer:1999mm}, which describes the 
influence of the quark transverse polarization on its momentum distribution inside 
an unpolarized parent hadron. Extraction of the latter is of great importance, 
because $h_1^\perp$ is believed to be responsible for the well known violation of 
the Lam-Tung sum rule~\cite{Falciano:1986wk,Guanziroli:1987rp,Conway:1989fs}, an 
anomalous big azimuthal asymmetry of the corresponding unpolarized Drell-Yan cross 
section, that neither NLO QCD calculations~\cite{Brandenburg:1993cj}, nor higher 
twists or factorization-breaking terms in NLO 
QCD~\cite{Brandenburg:1994wf,Eskola:1994py,Berger:1979du} are able to justify. 

In this paper, we will present numerical simulations for the unpolarized and
single-polarized Drell-Yan process $\bar{p} p^{\left( \uparrow \right)} 
\rightarrow \mu^+ \mu^- X$  at several center-of-mass energies using both configurations
where the antiproton beam hits a fixed proton target or it collides on another proton
beam. The goal is to estimate the minimum number of events needed to make 
the extraction of $h_1$ and $h_1^\perp$ feasible in future measurements at the 
HESR ring at GSI. In Sec.~\ref{sec:kin}, the kinematics and the
general cross section formulae are reviewed. In Sec.~\ref{sec:mc}, details about
the numerical simulation are given and results are presented in Sec.~\ref{sec:out}.
Finally, some remarks and conclusions are drawn in Sec.~\ref{sec:end}.

\section{The general framework}
\label{sec:kin}

In a Drell-Yan process, two (polarized) hadrons $H_1$ and $H_2$, carrying momentum 
$P_1$ and $P_2$ with $P_{1(2)}^2 = M_{1(2)}^2$ (and spin $S_1, S_2,$ such that 
$S_{1(2)}^2 = -1, \; P_{1(2)}\cdot S_{1(2)}=0$), annihilate into a lepton $l$ and an 
antilepton $\bar{l}$ with momenta $k_1$ and $k_2$ (with $k_{1(2)}^2 \sim 0$). The 
available initial squared energy in the center-of-mass (cm) frame is $s=(P_1+P_2)^2$. 
The time-like momentum transfer defines the hard scale and it is directly related to 
the invariant mass of the final state, i.e. $q^2 \equiv Q^2 = (k_1+k_2)^2 \geq 0$. 
Therefore, it is also indicated as $Q^2 \equiv M^2$; we will use both notations, in 
the following. The DIS regime is defined by the limit $Q^2, s \rightarrow \infty$, 
but keeping the ratio $0 \leq \tau = Q^2 / s \leq 1$ limited. In this regime, the 
Drell-Yan process can be factorized into an elementary annihilation process, 
convoluted with two soft functions describing the distribution of the two 
annihilating partons inside the corresponding hadrons. 

\begin{figure}
\centering
\includegraphics[width=7cm]{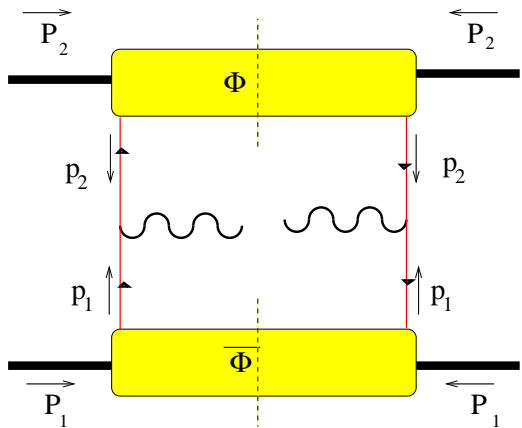}
\caption{The leading-twist contribution to the Drell-Yan process.}
\label{fig:handbag}
\end{figure}

If $M$ is constrained inside a range where the elementary annihilation can be safely 
assumed to proceed through a virtual photon converting into the final $l \, \bar{l}$ 
pair, while exploring the largest available range for $\tau$, the dominant 
contribution at leading order is depicted in 
Fig.~\ref{fig:handbag}~\cite{Ralston:1979ys}, where the blobs represent the 
correlation functions for the annihilating antiparton (labelled "1") and parton
(labelled "2"), respectively:
\bea
\bar{\Phi} (p_1; P_1,S_1) &= &\int \frac{d^4 z}{(2\pi )^4}\, e^{-i p_1 \cdot z}\, 
\langle P_1 S_1 | \psi(z) \, \bar{\psi}(0) |P_1, S_1 \rangle \; , \nn \\ 
\Phi (p_2; P_2,S_2) &= &\int \frac{d^4 z}{(2\pi )^4}\, e^{i p_2 \cdot z}\, \langle 
P_2, S_2 | \bar{\psi}(0) \, \psi(z) |P_2, S_2 \rangle \; .
\label{eq:phi}
\eea
When $Q^2 \rightarrow \infty$, the hard scale selects a light-cone dominant 
component for the hadron momenta, namely $P_1^+$ and $P_2^-$. Then, the parton 
momenta $p_{1(2)}$ are approximately aligned with the corresponding $P_{1(2)}$ and are
given by the light-cone fractions
\be
x_1 = \frac{p_1^+}{P_1^+} \simeq \frac{Q^2}{2\, P_1\cdot q} \; , \; 
x_2 = \frac{p_2^-}{P_2^-} \simeq \frac{Q^2}{2\, P_2\cdot q} \; , \;  
0\leq x_{1(2)} \leq 1 \; ,
\label{eq:lc-x}
\ee
while $q^+ = p_1^+$ and $q^- = p_2^-$, by momentum conservation~\cite{Boer:1999mm}. 

\begin{figure}
\centering
\includegraphics[width=7cm]{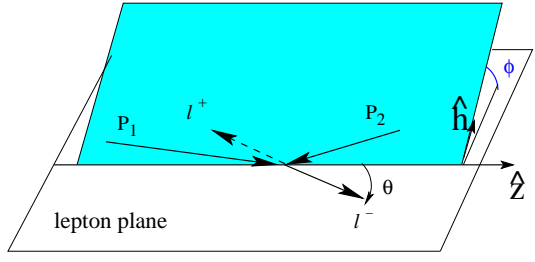}
\caption{The Collins-Soper frame.}
\label{fig:dyframe}
\end{figure}

As already anticipated in Sec.~\ref{sec:intro}, it is crucial to keep memory of the 
intrinsic transverse-momentum dependence inside the correlations functions of 
Eq.~(\ref{eq:phi}). Hence, the cross section
will be kept differential in ${\bf q}_{_T}$, which is bounded to ${\bf q}_{_T} = 
{\bf p}_{1_T} + {\bf p}_{2_T}$ by momentum conservation; ${\bf p}_{i_T}$ is the
transverse component of the parton momentum $p_i$ with respect to the axis defined by
the corresponding hadron 3-momentum ${\bf P}_i$, with $i=1,2$. In this context, if 
${\bf q}_{_T} \neq 0$ the annihilation direction is not known. Hence, it is 
convenient to select the socalled Collins-Soper frame~\cite{Collins:1977iv} 
(see fig.~\ref{fig:dyframe}), where
\bea
\hat{t} &= &\frac{q}{Q} \nn \\
\hat{z} &= &\frac{x_1 P_1}{Q} - \frac{x_2 P_2}{Q} \nn \\
\hat{h} &= &\frac{q_{_T}}{|{\bf q}_{_T}|} \; .
\label{eq:colsop-frame}
\eea

The azimuthal angles lie in the plane perpendicular to $\hat{t}$ and $\hat{z}$. In
particular, $\phi, \phi_{S_i},$ are the angles of $\hat{\bf h}, {\bf S}_{i_T},$ ($i=1,2$) 
with respect to the lepton plane, respectively, and 
$k_{i\perp}^\mu = (g^{\mu\nu} - \hat{t}^\mu \hat{t}^\nu +\hat{z}^\mu 
\hat{z}^\nu) k_{i\nu}$ the perpendicular component of $k_i^\mu$, with $i=1,2$.
Moreover, the cross section will be kept differential also in the solid angle for
the lepton production, i.e. $d\Omega = 2dy d\phi$, where the invariant 
$y=k_1^- / q^-$ reduces to $(1+\cos\theta)/2$ in the Collins-Soper frame, with 
$\theta$ defined in Fig.~\ref{fig:dyframe}. 

The hadronic tensor is obtained by taking the trace of the above correlation
functions, after integrating upon the corresponding suppressed light-cone
directions:
\be
W^{\mu\nu} = \frac{1}{3}\int dp_1^- dp_2^+\, \mbox{Tr} \left[ \bar{\Phi} (p_1;P_1,S_1)\, 
\gamma^\mu\,\Phi(p_2;P_2,S_2) \, \gamma^\nu \right] \Bigg\vert_{p_1^+=x_1
P_1^+,\,p_2^- = x_2 P_2^-} \, + \, \left( \begin{array}{ccc} q & \leftrightarrow &
-q \\ \mu & \leftrightarrow & \nu \end{array} \right) \; .
\label{eq:tensor}
\ee
Using the leading-twist parametrization for $\bar{\Phi}$ and $\Phi$ with explicit
dependence on ${\bf p}_{1_T}$ and ${\bf p}_{2_T}$~\cite{Boer:1998nt}, respectively, 
the fully differential cross section for the unpolarized Drell-Yan process 
$H_1 \, H_2 \rightarrow l\, \bar{l} \, X$  can be written as~\cite{Boer:1999mm}
\bea
\frac{d\sigma^o}{d\Omega dx_1 dx_2 d{\bf q}_{_T}} &= &\frac{\alpha^2}{3Q^2}\,
\sum_f\,e_f^2\,\Bigg\{ A(y) \, {\cal F}\left[ \bar{f}_1^f\, f_1^f \right] 
\nn \\
& & + B(y) \, \cos 2\phi \, {\cal F}\left[ \left( 2 \hat{\bf h}\cdot {\bf
p}_{1_T} \, \hat{\bf h} \cdot {\bf p}_{2_T} - {\bf p}_{1_T} \cdot {\bf p}_{2_T}
\right) \, \frac{\bar{h}_1^{\perp\,f}\,h_1^{\perp\,f}}{M_1\,M_2}\,\right]\, \Bigg\} 
\; ,
\label{eq:unpolxsect}
\eea
where $\alpha$ is the fine structure constant, $e_f$ is the charge of a parton with
flavor $f$, and
\bea
A(y) &= &\left( \frac{1}{2} - y + y^2 \right) \, \stackrel{\mbox{cm}}{=}\, 
\frac{1}{4}\left( 1 + \cos^2 \theta \right) \; , \nn \\
B(y) &= &y (1-y) \, \stackrel{\mbox{cm}}{=}\,\frac{1}{4}\, \sin^2 \theta \; .
\label{eq:lepton}
\eea
The functions $f_1^f, \, h_1^{\perp\,f},$ are the distributions of unpolarized 
and transversely polarized partons with flavor $f$ in unpolarized hadrons,
respectively. Both are convoluted with their antiparton partners according to the 
definition
\be
{\cal F} \left[ \bar{f}_1^f \, f_1^f \right] \equiv \int d{\bf p}_{1_T} d{\bf
p}_{2_T}\, \delta \left( {\bf p}_{1_T} + {\bf p}_{2_T} - {\bf q}_{_T} \right) \, 
\left[ \bar{f}_1^f (x_1,{\bf p}_{1_T})\, f_1^f(x_2,{\bf p}_{2_T}) + (1
\leftrightarrow 2) \right] \; .
\label{eq:convol}
\ee

The cross section~(\ref{eq:unpolxsect}) can be integrated upon the angular
distribution and the transverse momentum of the final lepton pair, 
\bea
\frac{d\sigma^o}{dx_1 dx_2} &= &\frac{1}{3}\, \sum_f \, e_f^2\, 
\frac{4\pi \alpha^2}{3 Q^2}\, \left[ \bar{f}_1^f(x_1) \, f_1^f(x_2) + 
(1\leftrightarrow 2)\right] \nn \\
&= &\frac{1}{3}\, \sum_f\, \int dQ^2\, \frac{d\hat{\sigma}^f}{dQ^2}\,\delta (\hat{s} 
- Q^2)\, \left[ \bar{f}_1^f(x_1) \, f_1^f(x_2) + (1\leftrightarrow 2)\right] \; ,
\label{eq:unpolxsect2}
\eea
where, if neglecting the parton masses, $\hat{s} \sim x_1 x_2 s$ is the available cm 
energy for the elementary annihilation, and the elementary QPM cross section has been 
put into evidence. At this point, it is useful to recall that $\tau = x_1 x_2$, and 
to introduce the new invariant $x_{_F} = x_1-x_2$, which is the "longitudinal" 
momentum of the annihilating partons in their cm frame with respect to the maximum 
"longitudinal" momentum available. If $d\sigma^o$ is considered differential also in 
the leptonic invariant mass $Q^2 \equiv M^2$, by transforming from the set 
$\{ x_1, x_2\}$ to the variables $\{\tau , x_{_F}\}$, and after integrating upon 
$d\tau$, we recover the well known result
\be
M^4\, \frac{d\sigma^o}{dM^2 dx_{_F}} = \frac{4\pi \alpha^2}{9}\,
\frac{\tau}{x_1+x_2}\, \sum_f \, e_f^2\, \left[ \bar{f}_1^f(x_1) \, f_1^f(x_2) + 
(1\leftrightarrow 2)\right]
\label{eq:unpolxsect3}
\ee
about the scaling of the cross section in the cm energy $s$, which has been
experimentally confirmed~\cite{Moreno:1990sf}. Interestingly, when considered 
differential in $\sqrt{\tau}$ and $x_{_F}$, the cross section falls like 
$\sim 1/s$ ; this remark will be reconsidered in the next Section because it 
influences the choice of the kinematical set-up.

When one of the annihilating hadrons is transversely polarized, i.e. for the
single-polarized Drell-Yan $H_1 \, H_2^\uparrow \rightarrow l\, \bar{l}\,X$, the
polarized part of the cross section becomes~\cite{Boer:1999mm}
\bea
& &\frac{d\Delta \sigma^\uparrow}{d\Omega dx_1 dx_2 d{\bf q}_{_T}} = 
\frac{\alpha^2}{3Q^2}\,\sum_f\,e_f^2\,|{\bf S}_{2_T}|\,\Bigg\{ A(y) \, 
\sin (\phi - \phi_{S_2})\, {\cal F}\left[ \hat{\bf h}\cdot {\bf p}_{2_T} \,
\frac{\bar{f}_1^f\, f_{1T}^{\perp\,f}}{M_2}\right] \nn \\
& &\quad - B(y) \, \sin  (\phi + \phi_{S_2})\, {\cal F}\left[  \hat{\bf h}\cdot 
{\bf p}_{1_T} \,\frac{\bar{h}_1^{\perp\,f}\, h_1^f}{M_1}\right] \nn \\
& &\quad  - B(y) \, \sin (3\phi - \phi_{S_2})\, {\cal F}\left[ \left( 
4 \hat{\bf h}\cdot {\bf p}_{1_T} \, (\hat{\bf h} \cdot {\bf p}_{2_T})^2 - 
2 \hat{\bf h} \cdot {\bf p}_{2_T} \, {\bf p}_{1_T} \cdot {\bf p}_{2_T} - 
\hat{\bf h}\cdot {\bf p}_{1_T} \, {\bf p}_{2_T}^2 \right) \, 
\frac{\bar{h}_1^{\perp\,f}\, h_{1T}^{\perp\,f}}{2 M_1\,M_2^2}\,\right]\, \Bigg\} 
\; ,
\label{eq:1polxsect}
\eea
where $f_{1T}^{\perp\,f}$ is the Sivers function, describing the distribution of
unpolarized partons with flavor $f$ in a transversely polarized hadron, while 
$h_{1T}^{\perp\,f}$ describes the distribution of transversely polarized
partons with flavor $f$ in transversely polarized hadrons. 

From Eq.~(\ref{eq:1polxsect}), a spin asymmetry weighted by $\sin (\phi -
\phi_{S_2})$ allows to isolate $f_{1T}^\perp$ through the known distribution $f_1$
with a mechanism similar to the Sivers effect in DIS with lepton 
beams~\cite{Sivers:1991fh}. Then, it could be possible to verify the interesting 
conjecture that the Sivers function should change sign when going from DIS to 
Drell-Yan~\cite{Collins:2002kn}, because of the different behaviour of the tower 
operator that must be inserted in Eq.~(\ref{eq:phi}) to link the space-time points 
$0$ and $z$ in order to make the whole correlator color-gauge invariant.

In Eq.~(\ref{eq:1polxsect}), another interesting spin asymmetry is obtained by using
the $\sin (\phi + \phi_{S_2})$ weight, because the transversity $h_1$ can be
extracted with a mechanism similar to the Collins effect~\cite{Collins:1993kk}. 
However, the convolution involves also the unknown chiral-odd function $h_1^\perp$, 
which luckily appears also in the leading-twist part of the unpolarized cross
section~(\ref{eq:unpolxsect}). A combined analysis of the $\cos 2\phi$ and $\sin
(\phi+\phi_{S_2})$ moments of azimuthal asymmetries for unpolarized and
single-polarized Drell-Yan cross sections, respectively, should allow a complete
determination of all the unknown distribution functions. The former asymmetry is
particularly interesting because it could represent a natural explanation of the
observed large azimuthal asymmetry of the cross section for the process $\pi^- \, A
\rightarrow \mu^+ \, \mu^- \, X$, where $A$ is either deuterium or tungsten, and 
the $\pi^-$ beam energy ranges from 140, 194, 286 
GeV~\cite{Falciano:1986wk,Guanziroli:1987rp} to 252 GeV~\cite{Conway:1989fs}. At leading
order in $\alpha_s$, the data can be parametrized as
\bea
\frac{1}{\sigma^o}\,\frac{d\sigma^o}{d\Omega} &\equiv &\left[ \frac{\alpha^2}{3Q^2}\,
\sum_f\,e_f^2 \int dx_1 dx_2 d{\bf q}_{_T}\,{\cal F}\left[\bar{f}_1^f \, f_1^f \right] 
\right]^{-1}\, \int dx_1 dx_2 d{\bf q}_{_T}\, \frac{d\sigma^o}{d\Omega dx_1 dx_2 d{\bf
q}_{_T}} \nn \\
&= &\frac{3}{4\pi}\,\frac{1}{\lambda +3}\, 
\left( 1+\lambda\,\cos^2\theta + \mu\,\sin^2\theta \cos \phi + 
\frac{\nu}{2}\,\sin^2\theta\cos 2\phi \right) \; ,
\label{eq:param}
\eea
with $\lambda \sim 1$ and $\mu \ll \nu \sim 30$
\%~\cite{Falciano:1986wk,Guanziroli:1987rp}. Both Leading Order (LO) and 
Next-to-Leading Order (NLO) perturbative QCD calculations give $\lambda \sim 1$, 
but $\mu \sim \nu \sim 0$~\cite{Brandenburg:1993cj}. Other mechanisms, like 
higher twists or factorization breaking terms at NLO, are not able to explain the 
size of $\nu$ in a consistent 
picture~\cite{Brandenburg:1994wf,Eskola:1994py,Berger:1979du}. On the contrary, 
from Eq.~(\ref{eq:unpolxsect}) it is evident that a leading-twist (hence, large) 
azimuthal $\cos 2\phi$ asymmetry in the unpolarized cross section is possible 
thanks to the distribution $h_1^\perp$~\cite{Boer:1999mm}. Since
the latter is defined to describe the transverse polarization of quarks in a
unpolarized hadron, it is intimately connected to the orbital motion of the parton
inside the hadron; the product $\bar{h}_1^\perp \, h_1^\perp$ should bring a change
of two units in the orbital angular momentum, such that the angular dependence
involves twice the azimuthal angle $\phi$.

In the next Section, we will see how to numerically simulate these azimuthal
asymmetries with a suitable Monte-Carlo.

\section{Monte-Carlo simulations}
\label{sec:mc}

In this Section we discuss a Monte-Carlo simulation for the specific processes 
$\bar{p} \, p \rightarrow \mu^+\, \mu^- \,X$ and 
$\bar{p} \, p^\uparrow \rightarrow \mu^+\, \mu^- \,X$.
The primary goal of the simulation is to estimate the number of events required for 
such a Drell-Yan experiment in order to access unambiguous information on the 
distribution functions of interest, namely the transversity $h_1$ and $h_1^\perp$.
Secondly, it is important also to clarify the role and the consequences of the 
unavoidable approximations and kinematical cuts. 

The cross sections for the unpolarized and polarized processes, Eqs.~(\ref{eq:unpolxsect})
and (\ref{eq:1polxsect}), respectively, are valid only in the framework of the QPM. In
order to implement QCD corrections (and to test the resulting formulae), we will heavily
refer to conventions and results of Ref.~\cite{Conway:1989fs}, where a thorough analysis
was performed on a large sample of muon pairs produced by a 252-GeV $\pi^-$ beam
interacting in a tungsten target. We will also use results from Ref.~\cite{Anassontzis:1987hk}
and Ref.~\cite{Boer:1999mm} in adapting the relations of Ref.~\cite{Conway:1989fs} to 
the case of an antiproton beam (with some care in matching the different notations) 
and of a polarized target, respectively. 

The Monte-Carlo events were generated from the following general expression for the cross
section:
\be
\frac{d\sigma}{d\Omega dx_{_F} d\tau d{\bf q}_{_T}} = K \, \frac{1}{s}\, 
A({\bf q}_{_T},x_{_F},M) \, F(x_{_F}, \tau) \, \sum_{i=1}^4 \, c_i (|{\bf
q}_{_T}|,x_{_F},\tau)\, S_i(\theta, \phi, \phi_{S_p})\; ,
\label{eq:mc-xsect}
\ee
or, equivalently,
\be
\frac{d\sigma}{d\Omega dx_{\bar{p}} dx_p d{\bf q}_{_T}} = K \, \frac{1}{s}\, 
A({\bf q}_{_T},x_{\bar{p}}-x_p,M)\, (x_{\bar{p}}+x_p) \, F'(x_{\bar{p}}, x_p)\, 
\sum_{i=1}^4\, c'_i (|{\bf q}_{_T}|,x_{\bar{p}},x_p) \, S_i(\theta, \phi, \phi_{S_p}) 
\; ,
\label{eq:mc-xsect1}
\ee
where the above mentioned scaling in $1/s$ has been put into evidence. As the reader can
argue from the formulae of previous Section, a (QPM-inspired) factorized structure 
for the cross section, but still differential in $d{\bf q}_{_T}$, implies that some 
simplifying assumption has been made on the ${\bf p}_{_T}$ dependence of the distribution 
functions, such that the convolution~(\ref{eq:convol}) could be solved. The actual
expression and its kinematical justification, as well as each other factor in the above 
equations, are separately discussed in the following.

\subsection{Kinematics and the transverse-momentum dependence}
\label{sec:pT}

For the handbag mechanism of Fig.~\ref{fig:handbag} to be the dominant contribution, 
some kinematical contraints must be fulfilled. First of all, the cm energy $s$ must be 
much bigger than the hadron masses involved. Secondly, in order to have a good control 
of the elementary annihilation process the invariant mass $M$ of the muon pair must not 
overlap with the hadronic resonance spectrum. The most natural choice is to select
$M$ such that 4 GeV $\leq M \leq 9$ GeV: in this range, between the $\psi'$ and
the first resonance of the $\Upsilon$ family, there should be no ambiguity in 
assuming that the elementary annihilation is followed by the creation of a virtual 
photon. 

We will consider both options where the antiproton beam hits a fixed proton target or it
collides on a proton beam. In the first case, if the target is at rest, we have 
\be
M^2 = \tau \, s = \tau \, (P_p + P_{\bar{p}})^2 = \tau\, 2 M_p (M_p + E_{\bar{p}}) \sim 
\tau \, 2 M_p E_{\bar{p}} \; ;
\label{eq:kin-s}
\ee
near the limit $\tau \rightarrow 1$ (i.e., assuming that all the available cm energy
flows into the final state), the upper limit for $M$ would require a maximum 
antiproton beam energy of roughly 40 GeV. With the lower cutoff of $M > 4$ GeV, this
corresponds to span the range $0.5 \lesssim \sqrt{\tau} \lesssim 1$, as discussed in the 
next Section and displayed in Fig.~\ref{fig:scatterplot}. Lower antiproton beam energies
would reduce this scatter plot to an unrealistic small portion of phase space limited to
very high values of both $x_p$ and $x_{\bar{p}}$, unless releasing the constraint on the
lower cutoff of the dilepton mass. In fact, for a beam energy of $E_{\bar{p}} = 15$ GeV
it is also possible to cover a wide range of $\tau$ with the constraint 1.5 GeV 
$\leq M \leq 2.5$ GeV, which grants no overlap with the positions of the $\phi$ 
and $J/\psi$ resonances. Therefore, in the following we will use both energies for the
antiproton beam. 

Recently, the discussions about the optimal setup configuration for spin asymmetry
measurements at the HESR at GSI have focussed also on the collider option where two
beams of protons and antiprotons collide at much higher $s$, particularly in the
socalled asymmetric mode with $E_p \neq E_{\bar{p}}$. Neglecting the hadron masses and
in the same conditions as in Eq.~(\ref{eq:kin-s}), we have
\be
M^2 \sim (P_p + P_{\bar{p}})^2 \sim  4 E_p E_{\bar{p}} \; ,
\label{eq:kin-s-coll}
\ee
because now $\hat{\bf P}_p = - \hat{\bf P}_{\bar{p}}$. Hence, for the case of an
antiproton beam energy of 15 GeV a proton beam of 3.3 GeV approximately gives $s \sim
200$ GeV$^2$. By keeping the above cutoffs for $M$, a large portion of the phase space
for $x_p$ and $x_{\bar{p}}$ can be explored, particularly at lower values. Moreover, at
larger energies the elementary process should be less affected by higher-order
corrections like subleading twists; hence, the theoretical description should be more
clean. Finally, as discussed below in Sec.~\ref{sec:target}, the collider mode offers
also the advantage that no dilution factors are introduced that affect the target
polarization. For all these reasons, in the following we will present simulations
for the single-polarized Drell-Yan process also at this kinematics.

The independent invariant fraction, $x_{_F} = x_{\bar{p}} - x_p$, can be 
redefined for practical purposes as~\cite{Conway:1989fs}
\be
x_{_F} \equiv \frac{2 q_{_L}}{\sqrt{s}} \; ,
\label{eq:newxF}
\ee
where $q_{_L}$ indicates the longitudinal component of the virtual photon momentum (in 
the following, we will also use the notation $q_{_T} = |{\bf q}_{_T}|$). The
definition~(\ref{eq:newxF}) puts in explicit evidence the content of $x_{_F}$ being the
fraction of the total available longitudinal momentum in the collision cm frame. In the 
DIS regime, it recovers the previous definition in terms of light-cone momentum 
fractions. It is important to note that the minimum and maximum $x_{_F}$ values depend on 
the energy. For the case of $E_{\bar{p}} = 40$ GeV and a fixed proton target, it turns 
out that $-0.7 \lesssim x_{_F} \lesssim 0.7$. 

A factorized transverse-momentum dependence in the differential cross section is
achievable assuming a simple parametrization for the ${\bf p}_{_T}$ dependence of the
distribution functions (typically, Gaussian-like), which allows to directly calculate the
convolution~(\ref{eq:convol}) and get a factorized product of two functions depending
separately upon ${\bf q}_{_T}$ and $x_{_F}, \tau$ (or, equivalently, $x_{\bar{p}}, 
x_p$ ; see, for example, the discussion of Sec.~V in Ref.~\cite{Boer:1999mm}). 
Alternatively, since 
the input ${\bf p}_{_T}$ dependence and, consequently, the obtained ${\bf q}_{_T}$ 
dependence are purely phenomenological, a parametrization for the latter can be assumed 
{\it ab initio} to fit the data. For example, the following form has been used in 
Sec.~V of Ref.~\cite{Conway:1989fs}, e.g.
\be
A(q_{_T},x_{_F},M) = \frac{5\,\displaystyle{\frac{a}{b}\,\left[ \frac{q_{_T}}{b}
\right]^{a-1}}}{\left[ 1 + \left(\displaystyle{\frac{q_{_T}}{b}}\right)^a \right]^6} \; ,
\label{eq:mcqT}
\ee
where $a(x_{_F},M),\,b(x_{_F},M),$ are parametric polynomials given in App. A of the same
reference. The normalization condition is 
\be
\int dq_{_T} \, A(q_{_T},x_{_F},M) = 1 \; .
\label{eq:mcqTnorm}
\ee
The function~(\ref{eq:mcqT}) gives a good reproduction of the observed $q_{_T}$
spectrum for $\pi - p$ Drell-Yan events (see fig.23 of Ref.~\cite{Conway:1989fs}).
However, in Ref.~\cite{Anassontzis:1987hk} the same analysis has been repeated also for 
$\bar{p} - p$ events and the two distributions are similar for $q_{_T} \gtrsim 1$ 
GeV/$c$, while above 3 GeV/$c$ big error bars prevent from any comparison. Therefore,
we have adopted the same distribution~(\ref{eq:mcqT}) also in our analysis. 

With the parametrization displayed in App. A of Ref.~\cite{Conway:1989fs}, $a$ and $b$ 
turn out to be smoothly dependent on $x_{_F}$ and $M$, while the calculated 
$\langle {\bf q}_{_T}^2\rangle$ is a smooth function only in $M$ (after integrating on 
the whole range of $x_{_F}$): it displays a strong decrease in $x_{_F}$ for 
$x_{_F}\rightarrow 0.7$ after averaging on the whole range of $M$. The smooth dependence 
of $\langle {\bf q}_{_T}^2\rangle$ on $M$ reflects in its experimentally observed scale 
dependence $\sim \sqrt{s}$, but if the cut-offs in the invariant mass are the same for 
two experiments at different energies (as it is the case between the measurement of
Ref.~\cite{Conway:1989fs} and the present simulation), the average squared transverse 
momentum results approximately the same. 

We have cut the $q_{_T}$ distribution from below, because events at low $q_{_T}$ do not 
contribute to the spin asymmetry, but rather to the background, and they would represent 
a sort of artificial dilution factor. However, since this cutoff has drastic consequences 
on the final number of available events, a good compromise at these energies is 
represented by $q_{_T} > 1$ GeV/$c$, in accordance with several previous experimental 
observations. Such a constraint implies a reduction of the total number of available 
events by approximately 50\%.

\subsection{QCD corrections and the partonic-momentum dependence}
\label{sec:Kx1x2}

The well established experimental observation that Drell-Yan pairs are distributed with 
$\langle q_{_T}\rangle > 1$ GeV/$c$ and depending on $\sqrt{s}$, suggests that
sizeable QCD corrections are needed on top of the QPM, because confinement alone induces 
much smaller quark intrinsic transverse momenta. The Feynman diagrams typically involve $q
\bar{q}$ annihilations into gluons or quark-gluon scattering (see Fig.~3 of 
Ref.~\cite{Conway:1989fs}). They have been calculated introducing two main levels of
approximation~\cite{Altarelli:1979ub,Berger:1982xr}. The first one is the so-called 
Leading-Log 
Approximation (LLA), where the leading logarithmic corrections to the Drell-Yan cross 
section can be resummed at any order in the strong coupling constant $\alpha_s$, with the 
final net effect that the parton distribution functions get an additional scale 
dependence on $M^2$. Following the prescriptions of Ref.~\cite{Buras:1977yj}, the DGLAP 
evolution could be obtained by describing the functions with parameters explicitly 
depending on $\log M^2$ (see Apps.~A,B and D in Ref.~\cite{Conway:1989fs} for further 
details). However, since the $M$ range explored in Ref.~\cite{Conway:1989fs} is 
practically the same assumed here, also the same parametrization described there in 
App.~A is here retained; correspondingly, the function $F'$ in Eq.~(\ref{eq:mc-xsect1}) 
does no longer depend on the scale $M^2$. We have explicitly
verified that starting from the parametrization of Ref.~\cite{Conway:1989fs} (and after a
suitable transformation due to different conventions), our Monte-Carlo gives a good 
approximation of the results of Ref.~\cite{Anassontzis:1987hk} with an antiproton beam of
125 GeV.

\begin{figure}
\centering
\includegraphics[width=7cm]{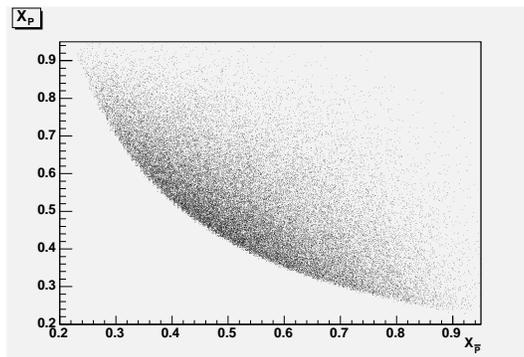}
\caption{The scatter plot for 48000 events of Drell-Yan muon pairs produced by the
collision of a 40 GeV antiproton beam on a proton target in the kinematic conditions 
discussed in the text.}
\label{fig:scatterplot}
\end{figure}

The function $F'$ in Eq.~(\ref{eq:mc-xsect1}) is given by
\be
F'(x_{\bar{p}},x_p) = \frac{\alpha^2}{3Q^2}\,\sum_f\,e_f^2\,\bar{f}_1^f(x_{\bar{p}})
\, f_1^f (x_p) + (\bar{p} \leftrightarrow p) \; , 
\label{eq:mcF}
\ee
namely it is the symmetric part of the unpolarized cross section that has been factorized
out for convenience (see Eq.~(\ref{eq:unpolxsect})). With $f_1(x)$ parametrized as 
in App.~A of Ref.~\cite{Conway:1989fs} for the various flavors $f=u,d,s$, we show in
Fig.~\ref{fig:scatterplot} the scatter plot in $x_{\bar{p}}$ and $x_p$ of some 
48000 events sorted by the cross section of Eq.~(\ref{eq:mc-xsect1}) including 
Eq.~(\ref{eq:mcF}) and assuming the kinematics corresponding to $E_{\bar{p}} = 40$ GeV 
with the cutoffs $q_{_T} > 1$ GeV/$c$, $4 \leq M \leq 9$ GeV and 
$-0.7 \lesssim x_{_F} \lesssim 0.7$. 

\begin{figure}
\centering
\includegraphics[width=7cm]{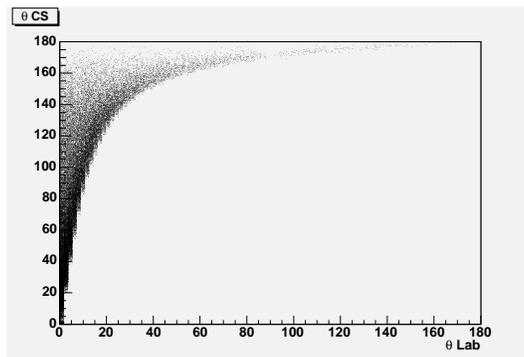}
\caption{The correlation between the polar angle of $\mu^+$ in the laboratory,
$\theta_{\rm lab}$, and in the Collins-Soper frame, $\theta$, for the collision of
antiprotons with 40 GeV energy and fixed proton targets.}
\label{fig:framecorr}
\end{figure}

The line bisecting the plot at 45 deg. corresponds to $x_{_F}=0$; parallel lines above 
and below it indicate $x_{_F}>0$ and $x_{_F}<0$, respectively. Hyperboles with 
$x_{\bar{p}} x_p =$ const. select various values of $\tau$: moving to the upper
right corner of the figure corresponds to the limit $\tau \rightarrow 1$, where events are
rare; on the contrary, approaching the lower cutoff $\sqrt{\tau} \sim 0.5$ the 
distribution becomes much more dense. Nevertheless, the projection along each axis 
displays a significant number of events also at the boundary values of $x_{\bar{p}}, x_p$. 
They correspond to large positive and negative values of $x_{_F}$, which in turn 
indicate, in the Collins-Soper frame, small and large values of the muon pair polar 
angle $\theta$, respectively. For a fixed target and in the laboratory frame, since we 
have antipartons in the beam and partons in the target, large and positive values of 
$x_{_F}$ correspond to "forward" events where, by convention, the polar angle 
$\theta_{\rm lab}$ of $\mu^+$ is small. Viceversa, large negative values of $x_{_F}$ 
correspond to "backward" events with large $\theta_{\rm lab}$. This is at variance with 
the experiments of Refs.~\cite{Conway:1989fs,Anassontzis:1987hk}, where events with only 
large positive $x_{_F}$ were collected because of the higher beam energy and of the 
stronger absorption of the unpolarized target. In Fig.~\ref{fig:framecorr}, the 
correlation between $\theta$ and $\theta_{\rm lab}$ is shown in order to estimate the 
possible geometry of the apparatus for detecting the muons. The points are spread because 
the relation between the two frames depends on $x_{\bar{p}}, x_p,$ and $q_{_T}$.

The second level of approximation in the QCD corrections to the QPM result is referred to
as Next-to-Leading-Log Approximation (NLLA). It amounts to include in the calculation all
processes at first order in $\alpha_s$ involving a quark, an antiquark and a 
gluon~\cite{Altarelli:1979ub,Berger:1982xr}. The effect is large, because the cross 
section has approximately twice the size of the corresponding one in the QPM. This 
correction turns out to be roughly independent of $x_{_F}$ and $M^2$ (except for the 
kinematical upper limits) and it is usually indicated as the $K$ factor. As
explained in Ref.~\cite{Anassontzis:1987hk}, $K$ depends on the choice of the
parametrization of the distribution functions through their normalization, but it also 
grows like $\sqrt{\tau}$~\cite{Conway:1989fs}. However, in the following we will consider 
azimuthal asymmetries, which are defined as ratios of cross sections. Hence, it is 
legitimate to expect that such corrections in the numerator and in the denominator 
approximately compensate each other. Indeed, for fully polarized Drell-Yan 
reactions at high energy the spin asymmetry is weakly
affected by QCD corrections~\cite{Martin:1998rz}. In any case, for the 
considered antiproton energy $E_{\bar{p}} = 40$ GeV at $\sqrt{\tau} \sim 0.5$, where most 
of the events are concentrated, the best compromise is to assume the constant value 
$K=2.5$.

\subsection{The angular distribution}
\label{sec:angles}

If the orientation of the $\hat{x}$ axis of the Collins-Soper frame (pointing along 
${\bf q}_{_T}$) with respect to the laboratory frame is indicated with $\phi^l$, then
it turns out that $\phi^l = - \phi_{S_p}$ because, in practice, the latter angle 
gives the position of the $\hat{x}$ axis of the laboratory frame (pointing along 
${\bf S}_{p_T}$) in the Collins-Soper frame. The whole solid angle $(\theta, \phi)$ of the 
final muon pair in the Collins-Soper frame is randomly distributed in each variable. 
The explicit form for sorting the angular distribution in the Monte-Carlo is
\bea
\sum_{i=1}^4\, c'_i (q_{_T},x_{\bar{p}},x_p) \, S_i(\theta, \phi, \phi_{S_p}) &= 
&1 + \cos^2 \theta \nn \\
& &+ \frac{\nu (x_{\bar{p}},x_p,q_{_T})}{2}\, \sin^2\theta \, \cos 2\phi \nn \\
& &+ |{\bf S}_{p_T}|\, c_4 (q_{_T},x_{\bar{p}},x_p)\, S_4 (\theta, \phi, \phi_{S_p}) \; ,
\label{eq:mcS}
\eea
which corresponds to the parametrization~(\ref{eq:param}) for $\lambda = 1, \mu = 0,$ as
suggested by experimental observations, and includes also a polarized term as suggested
by Eq.~(\ref{eq:1polxsect}). 

If quarks were massless, the virtual photon would be only transversely polarized and the
angular dependence would be described by the functions $c_1 = S_1 = 1$ and $c_2 = 1, \,
S_2 = \cos^2 \theta$. Violations of such azimuthal symmetry induced by the function 
$c_3 \equiv \textstyle{\frac{\nu}{2}}$ are due to the longitudinal polarization of 
the virtual photon and to the fact that quarks have an intrinsic transverse momentum 
distribution. QCD corrections influence such function, which in principle depends 
also on $x_{\bar{p}}, x_p,$ and $M^2$ (see App.~A of Ref.~\cite{Conway:1989fs}). In 
practice, we follow Ref.~\cite{Boer:1999mm} (where $\nu = 2 \kappa$ and $\kappa$ is 
given in Sec.~V) and assume the simple form
\be
\nu (q_{_T}) = 3.5 \, \frac{4 M_{_C}^2 q_{_T}^2}{(4 M_{_C}^2 + q_{_T}^2)^2} \; ,
\label{eq:mcnu}
\ee
where $M_{_C} = 2.3$ GeV. 

The last term in Eq.~(\ref{eq:mcS}) corresponds to the polarized part of the cross
section, which is fully described in Eq.~(\ref{eq:1polxsect}). Since we are interested
in the convolution of $h_1$ and $h_1^\perp$, we assume that the angular dependence of
the azimuthal spin asymmetry is given by 
\be
S_4 (\theta, \phi, \phi_{S_p}) = \sin^2 \theta \, \sin (\phi + \phi_{S_p}) \; . 
\label{eq:mcS4}
\ee
Recalling that in Eq.~(\ref{eq:mc-xsect1}) the azimuthally symmetric unpolarized part 
$F'(x_{\bar{p}},x_p)$ of the cross section has been factorized out, the 
corresponding coefficient $c_4$ in Eq.~(\ref{eq:mcS}) becomes
\be
c_4 (q_{_T},x_{\bar{p}},x_p) = - \frac{ \sum_f\,e_f^2\,{\cal F}\left[  \hat{\bf h}
\cdot {\bf p}_{\bar{p}_T} \,\displaystyle{\frac{\bar{h}_1^{\perp\,f}\,
h_1^f}{M_{\bar{p}}}}\right]}{\sum_f\,e_f^2\,
{\cal F}\left[ \bar{f}_1^f\, f_1^f \right]} \; .
\label{eq:mcc4}
\ee
Following Ref.~\cite{Boer:1999mm}, we introduce some model assumptions on the behaviour
of the unknown $h_1^\perp (x,{\bf p}_{_T})$. Its ${\bf p}_{_T}$ dependence is of the
Gaussian type, while the $x$ dependence is given directly by $f_1(x)$: 
\be
h_1^{\perp\,f}(x,{\bf p}_{_T}^2) = \frac{\alpha_{_T}}{\pi}\,
\frac{M_{_C}}{{\bf p}_{_T}^2 +M_{_C}^2} \, e^{-\alpha_{_T} {\bf p}_{_T}^2}\, f_1^f (x) 
\; ,
\ee
where $\alpha_{_T} = 1$ GeV$^{-2}$. Then, the $c_4$ coefficient turns out to be 
(see Sec.~VI of Ref.~\cite{Boer:1999mm} for all details):
\be
c_4 (q_{_T},x_{\bar{p}},x_p) = - \frac{2 q_{_T} M_{_C}}{4 M_{_C}^2 + q_{_T}^2} \, 
\frac{\sum_f\, e_f^2\, \bar{f}_1^f (x_{\bar{p}}) \,  h_1^f (x_p) }
{ \sum_f\, e_f^2\, \bar{f}_1^f (x_{\bar{p}})  \, f_1^f (x_p) } \; .
\ee
For sake of simplicity, we assume that we can approximate the contribution of each 
flavor to the parton distributions by a corresponding average function, such that we
finally get
\be
c_4 (q_{_T},x_{\bar{p}},x_p) \sim - \frac{2 q_{_T} M_{_C}}{4 M_{_C}^2 + q_{_T}^2} \, 
\frac{\langle \bar{f} (x_{\bar{p}}) \rangle \, \langle h_1 (x_p) \rangle}
{\langle \bar{f} (x_{\bar{p}}) \rangle \, \langle f_1 (x_p) \rangle} \equiv 
- \frac{2 q_{_T} M_{_C}}{4 M_{_C}^2 + q_{_T}^2} \, 
\frac{\langle h_1 (x_p) \rangle}{\langle f_1 (x_p) \rangle} \; .
\label{eq:mcc4-1}
\ee

In order to disentangle the various azimuthal parts in the cross section, the most
straightforward way is to integrate the corresponding azimuthal asymmetry with a proper
weight depending on $\phi$. However, if the $\sin (3\phi - \phi_{S_2})$ contribution to
Eq.~(\ref{eq:1polxsect}) is assumed small, a simpler procedure is available. In fact,
for the remaining leading-twist contributions
\bea
P_1 (\phi) &= &\cos 2\phi \; , \nn\\
P_2(\phi) &= &\sin (\phi - \phi_{S_p}) \; , \nn \\
P_3(\phi) &= &\sin (\phi + \phi_{S_p}) \; , 
\label{eq:allphi}
\eea
it is possible to show that for each bin $x_p$ the asymmetry between the subsets of 
positive and negative values of $P_i (\phi)$ (spanning the whole $\phi$ range) does not 
depend on $P_j$ for $j \neq i$. In the following Section, results will be presented and 
commented for azimuthal asymmetries constructed in this way for $P_1 (\phi)$ and 
$P_3 (\phi)$. Two choices with opposite features will be selected for the ratio 
$\langle  h_1 (x_p) \rangle / \langle  f_1 (x_p) \rangle$, namely the ascending 
function $2x_p$ and the descending one $2 (1-x_p)$. The goal is to determine the 
minimum number of events (compatible with the kinematical setup and cutoffs) required 
to produce azimuthal asymmetries that can be clearly distinguished like the 
corresponding originating distributions. In fact, this would be equivalent to state that
some analytic information on $h_1(x)$ could be extracted from such spin asymmetry 
measurements.

\subsection{Polarized target and dilution factor}
\label{sec:target}

When the transversely polarized proton is considered as a fixed target, in reality we 
assume a $NH_3$ target with 85\% transverse polarization for the $H$ nucleus. In the 
ideal conditions of scattering from a black sphere, the $N$ hadronic cross section is 
exactly ten times bigger than the one for $H$, i.e. 390 mb~\cite{Segre:??}. This means 
that, from the point of view of hadronic collisions, the $N$ nucleus is equivalent to 5 
protons and 5 neutrons with no difference between protons and neutrons. In reality, 
this picture can be valid only at very small $x_{\bar{p}}$ and $x_p$, where the valence 
quarks do not dominate. However, as it is clear from Fig.~\ref{fig:scatterplot} this 
domain is not accessible to our simulation. Rather, the events are concentrated at 
$x_i > 0.1, \, i=\bar{p},p$, where the valence distributions are dominant and the 
annihilation occurs between a quark and an antiquark of the same flavor. Hence, the 
antiproton of the beam "prefers" to hit protons rather than neutrons. A convenient 
estimate would be to assume that in the $NH_3$ target the $N$ nucleus behaves like if 
it was made of 5 protons and 4 neutrons, such that out of 12 events, 9 come from 
collisions inside $N$ and 3 inside $H_3$. Therefore, in the scatter plot of 
Fig.~\ref{fig:scatterplot} with 48000 events, only 12000 are related to $H_3$ with 
85\% of transverse polarization, while 36000 are produced on the unpolarized $N$ 
nucleus. The Monte-Carlo simulation takes into account such 
a combination of dilution factors.  

When the $\bar{p} \, p^\uparrow \rightarrow \mu^+\, \mu^- \,X$ process is considered as 
the collision of two beams, a systematic dilution factor 0.85 is applied to the 
proton polarization. The statistics of the simulation greatly benefits from this weaker
dilution and, in fact, a much smaller sample of events is needed to get reasonable
results, as it will become evident in the next Section.

\section{Results}
\label{sec:out}

In this Section, we present results of a Monte-Carlo simulation for the 
$\bar{p} \, p^\uparrow \rightarrow \mu^+\, \mu^- \,X$ and 
$\bar{p} \, p \rightarrow \mu^+\, \mu^- \,X$ processes in order to make realistic 
estimates of the minimum number of events required to extract as detailed information as
possible on the chiral-odd distributions $h_1$ and $h_1^\perp$. First, we consider an 
antiproton beam of 40 GeV energy hitting a fixed proton target and producing muon pairs 
with invariant mass in the range 4 GeV $\leq M \leq 9$ GeV and total transverse 
momentum $q_{_T} > 1$ GeV/$c$. Then, we will select the lower antiproton beam energy of 
15 GeV with the dilepton invariant mass in the range 1.5 GeV $\leq M \leq 2.5$ GeV and we 
will explore both possibilities of a fixed proton target and of a collision with a proton 
beam of 3.3 GeV energy, such that the cm available squared energy is $s
\sim 200$ GeV$^2$ (see Sec.~\ref{sec:pT}). For fixed targets at such low energy, the 
QPM picture leading to Eq.~(\ref{eq:mc-xsect}), or (\ref{eq:mc-xsect1}), is not well 
justified because of the unavoidable corrections due to higher-twist contributions. But 
we stress that the primary goal of this simulation is to explore the feasibility of the 
extraction of the $h_1$ and $h_1^\perp$ distributions, not to make a precision 
calculation. Hence, we will retain the same approach in all kinematical cases above
described. 

The transversely polarized proton target is obtained from a $NH_3$ molecule where each 
$H$ nucleus is 85\% transversely polarized and the number of "polarized" collisions with 
each $H$ nucleus is 25\% of the total number of collisions. The events are sorted 
according to the cross section of Eq.~(\ref{eq:mc-xsect1}) (reduced by the above 
dilution factor) and supplemented by Eqs.~(\ref{eq:mcqT}) and 
(\ref{eq:mcF})-(\ref{eq:mcc4-1}). In the following, we will study the azimuthal
asymmetries generated by the $\cos 2\phi$ and $\sin (\phi + \phi_{S_p})$ dependences in
the cross section (see Eqs.~(\ref{eq:unpolxsect}), (\ref{eq:1polxsect}), (\ref{eq:mcS})
and (\ref{eq:mcS4})). For each $x_p$ bin, the asymmetry is constructed by dividing the
sample in two groups, one for positive values in the $\phi$ dependence ($U$) and another
one for negative values ($D$), and taking the ratio $(U-D)/(U+D)$; the azimuthal $\phi$
dependence is considered in the Collins-Soper frame. 

We have initially taken samples of 100000 and 20000 events for fixed target 
and collider modes, respectively. The above cuts in $M$ and $q_{_T}$ are implemented also
by a restriction on the $\theta$ angular dependence to the range 60 deg. $< \theta < 
120$ deg., where events are mostly concentrated; outside these limits, in fact, the
azimuthal asymmetries are too small. Analogously, as a general rule $x_p$ bins with less 
than 100 events are considered empty. All these cuts reduce the starting samples 
approximately by a factor 2.5; conventionally, we will indicate the surviving 40000 and 
8000 events, respectively, as "good" events. Statistical errors for the above mentioned 
azimuthal asymmetry $(U-D)/(U+D)$ are obtained by making 20 independent repetitions of the 
simulation for each considered case, and then calculating for each $x_p$ bin the average 
asymmetry value and the variance. We checked that 20 repetitions are a
reasonable threshold to have stable numbers, since the results do not change significantly
when increasing the repetitions from 6 to 20.

\begin{table}
\caption{\label{tab:totxsect} Total absorption cross sections per nucleon for 
antiprotons at the indicated energies and for various invariant masses of the 
Drell-Yan pair (see text for a discussion of the kinematics and the cutoffs).}
\begin{ruledtabular}
\begin{tabular}{ccccc}
$E_{\bar{p}}$ (GeV) & mode & event sample & $M$ (GeV) & $\sigma_{tot}$ (nb/nucleon) \\
\hline
40 & fixed target & 40000 & $4 \div 9$ & 0.02 \\
40 & fixed target & 40000 & $1.5 \div 2.5$ & 1.6 \\
15 & fixed target & 40000 & 4 & $4 \times 10^{-4}$ \\
15 & fixed target & 40000 & $1.5 \div 2.5$ & 0.8 \\
15 & collider & 8000 & $4 \div 9$ & 0.1 \\
15 & collider & 8000 & $1.5 \div 2.5$ & 2.4 \\
\end{tabular}
\end{ruledtabular}
\end{table}

From our Monte-Carlo results, we deduce a total cross section $\sigma_{tot} = 0.02$ nb 
per single nucleon for the absorption of antiprotons in a 40 GeV beam hitting a 
polarized $NH_3$ target in the above discussed conditions and producing final lepton 
pairs with invariant mass $M \geq 4$ GeV. This cross section is $2 \times 10^9$ 
times smaller than the total nucleon absorption cross section per single nucleon at the
considered energy and under the conditions for a complete absorption, since the latter 
approximately amounts to 40 mb. In other words, when the flux of
actually absorbed antiprotons is deduced from the experimental setup, the above cross 
section states that one out of $2 \times 10^9$ such antiprotons produces a "good" event 
for the selected kinematics. In Tab.~\ref{tab:totxsect}, we list the total cross sections 
for all the explored kinematics and for various ranges of invariant masses. If ${\cal L}$ 
is the luminosity in units (cm$^{-2}$s$^{-1}$), the product $\sigma_{tot} {\cal L}$ gives 
the number of "good" events per nucleon and per second. For example, we deduce that in 
the collider mode for $1.5\leq M \leq 2.5$ GeV we would approximately get 60000 
"good" events/month with ${\cal L} = 10^{31}$ (cm$^{-2}$s$^{-1}$).

\begin{figure}[h]
\centering
\includegraphics[width=10cm]{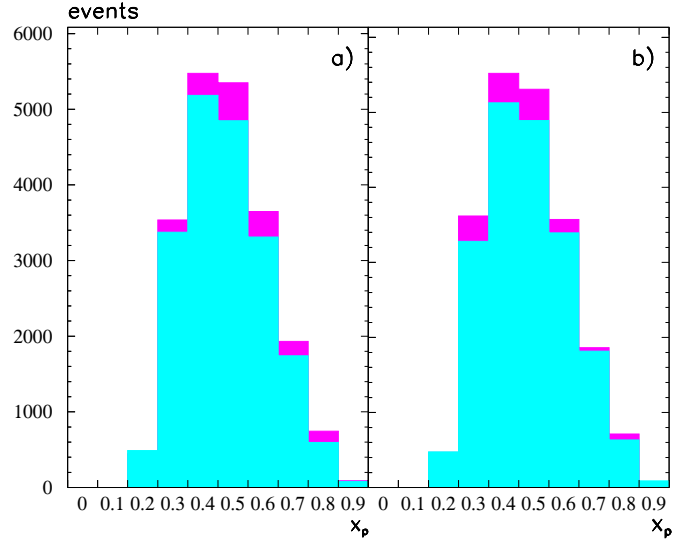}
\caption{The sample of 40000 events for the 
$\bar{p} \, p^\uparrow \rightarrow \mu^+\, \mu^- \,X$ process on a fixed proton target
with an antiproton beam energy $E_{\bar{p}} = 40$ GeV and with a muon pair of 
invariant mass $4 \leq M \leq 9$ GeV and transverse momentum $q_{_T} > 1$
GeV/$c$ (for further details on the cutoffs, see text). a) Left panel for the choice 
$\langle h_1(x_p) \rangle / \langle f_1(x_p) \rangle = 2 x_p$ (brackets mean that 
each flavor contribution in the numerator is replaced by a common average term,
similarly in the denominator; for further details, see text). b) Right panel for 
$\langle h_1(x_p) \rangle / \langle f_1(x_p) \rangle = 2 (1-x_p)$. For each bin, 
the darker histogram corresponds to positive values of $\sin(\phi + \phi_{S_p})$ in
Eqs.~(\protect{\ref{eq:mcS}}) and (\protect{\ref{eq:mcS4}}), the superimposed lighter 
one to negative values.}
\label{fig:hst-40}
\end{figure}

\subsection{Polarized Drell-Yan}
\label{sec:polDY}

In Fig.~\ref{fig:hst-40}, the sample of 40000 "good" events for the 
$\bar{p} \, p^\uparrow \rightarrow \mu^+\, \mu^- \,X$ process is displayed for the 
antiproton beam energy $E_{\bar{p}} = 40$ GeV and for the case of a fixed proton
target. The left panel corresponds to the choice 
$\langle h_1(x_p) \rangle / \langle f_1(x_p) \rangle = 2 x_p$, while the right one to
the function $2 (1 - x_p)$, according to Eq.~(\ref{eq:mcc4-1}). Results are reported 
in $x_p$ bins over the entire range, but the bins at the boundaries, corresponding to 
$x_p < 0.2$ and $x_p > 0.8$, are scarcely or not at all populated, according to the phase
space distribution of Fig.~\ref{fig:scatterplot}. For each bin, two groups of events are 
stored: one corresponding to positive values of $\sin (\phi + \phi_{S_p})$ in
Eqs.~(\ref{eq:mcS}) and (\ref{eq:mcS4}) (represented by the darker histogram), the 
other one corresponding to negative values (represented by the superimposed lighter 
histogram).

\begin{figure}[h]
\centering
\includegraphics[width=10cm]{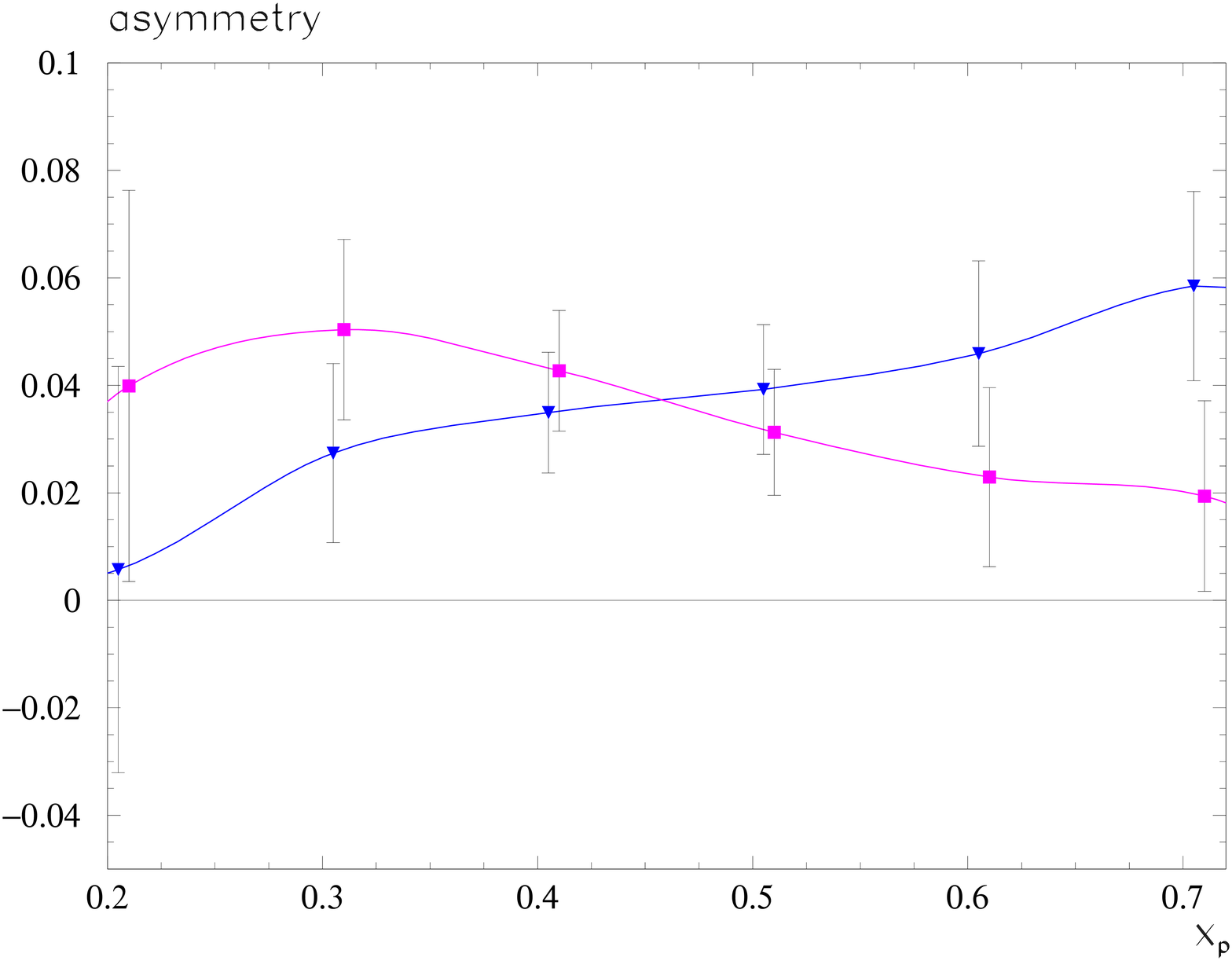}
\caption{Azimuthal asymmetry $(U-D)/(U+D)$ between cross sections in the previous figure
corresponding to the darker histograms ($U$) and superimposed lighter histograms ($D$), 
as bins in $x_p$. Downward triangles for the case when  
$\langle h_1(x_p) \rangle / \langle f_1(x_p) \rangle = 2 x_p$, squares when it equals 
$2 (1-x_p)$. Continuous lines are drawn to guide the eye. Error bars due to statistical 
errors only, obtained by 20 independent repetitions of the simulation (see text for further
details).}
\label{fig:as-40}
\end{figure}

In Fig.~\ref{fig:as-40}, the resulting asymmetry $(U-D)/(U+D)$, between cross
sections with positive ($U$) and negative ($D$) values of $\sin (\phi + \phi_{S_p})$, 
is plotted again in $x_p$ bins for both choices: downward triangles for the case 
when $\langle h_1(x_p) \rangle / \langle f_1(x_p) \rangle = 2 x_p$, squares when it 
equals $2 (1-x_p)$. The error bars represent statistical errors only.

\begin{figure}[h]
\centering
\includegraphics[width=10cm]{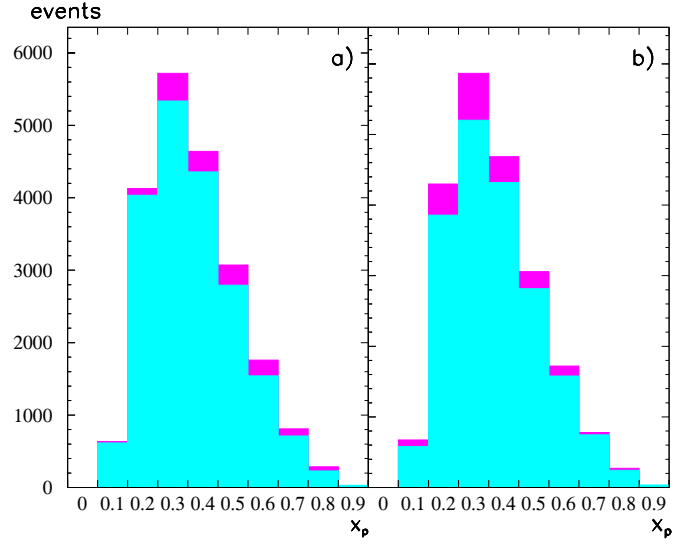}
\caption{Same as in Fig.~\protect{\ref{fig:hst-40}} but with the antiproton beam energy 
$E_{\bar{p}} = 15$ GeV and with a muon pair of invariant mass 
$1.5 \leq M \leq 2.5$ GeV.}
\label{fig:hst-15}
\end{figure}

Here and in the following cases, both functional forms for 
$\langle h_1(x_p) \rangle / \langle f_1(x_p) \rangle$, once integrated in $x_p$ over the 
entire range, give the same overall size 1. Extrapolating information from 
other sources (lattice calculations, other experiments like semi-inclusive DIS, 
theoretical properties like the Soffer bound, etc..) suggests that this size may be a 
reasonable expectation~\cite{Boer:1999mm}. From Fig.~\ref{fig:as-40}, we first conclude 
that 40000 "good" events of antiprotons with energy 40 GeV hitting fixed protons and
producing muon pairs with $4 \leq M \leq 9$ GeV, are enough to produce a
statistically nonvanishing azimuthal asymmetry of about 5\% on the average. But it is
difficult to disentangle the two different functional forms of the ratio except for 
very high $x_p \gtrsim 0.6$, where the QPM picture becomes questionable. 
Equivalently, it is difficult to extract unambiguous information on the transversity 
distribution $h_1(x)$. Moreover, from Tab.~\ref{tab:totxsect} we deduce that a luminosity 
${\cal L} = 10^{31}$ cm$^{-2}$s$^{-1}$ would give a rate of 500 "good" events/month. The 
same conclusion holds if the absolute size of the ratio is expected in the range 
$0.5 \div 1$. Only a smaller number by an order of magnitude would imply a much larger 
sample of events in order to reach the same sensitivity.

\begin{figure}[t]
\centering
\includegraphics[width=10cm]{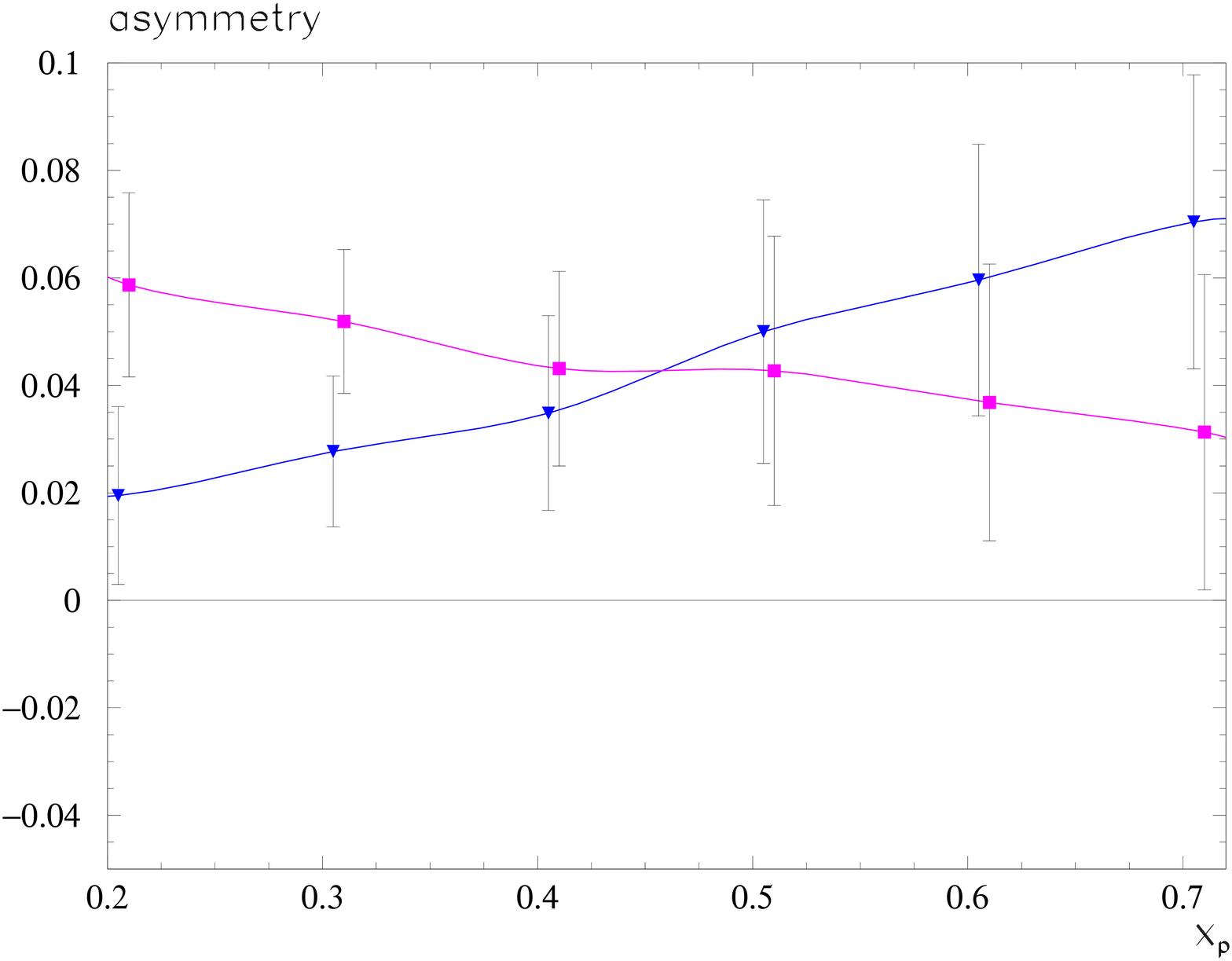}
\caption{Azimuthal asymmetry collected in $x_p$ bins in the same kinematics as in
Fig.~\protect{\ref{fig:hst-15}} and with the same notations as in
Fig.~\protect{\ref{fig:as-40}}. Continuous lines are drawn to guide the eye. Error bars
from statistical errors only.}
\label{fig:as-15}
\end{figure}

The situation is somewhat better at the lower antiproton beam energy
$E_{\bar{p}} = 15$ GeV, because in the scatter plot (analogous of 
Fig.~\ref{fig:scatterplot}) the lowest cutoff in $\tau = x_p x_{\bar{p}} = M^2/s$ is 
smaller since now $1.5 \leq M \leq 2.5$ GeV.
The highest density of events (and the best statistics) is now reached at lower values of 
$x_p$. In fact, while in Fig.~\ref{fig:hst-15} we show the usual sample of 40000 "good"
events in the same conditions and notations as in Fig.~\ref{fig:hst-40} but for
$E_{\bar{p}} = 15$ GeV and $1.5 \leq M \leq 2.5$ GeV, in Fig.~\ref{fig:as-15} the
resulting azimuthal asymmetry (again, with the same notations as in Fig.~\ref{fig:as-40})
clearly shows how the ascending (descending) behaviour of 
$\langle h_1(x_p) \rangle / \langle f_1(x_p) \rangle = 2 x_p \, [2(1-x_p)]$ reflects in
the corresponding asymmetry making the two results statistically distinguishable for $x_p
\lesssim 0.4$ (and, approximately, also for very high values $x_p \gtrsim 0.7$).

\begin{figure}[h]
\centering
\includegraphics[width=10cm]{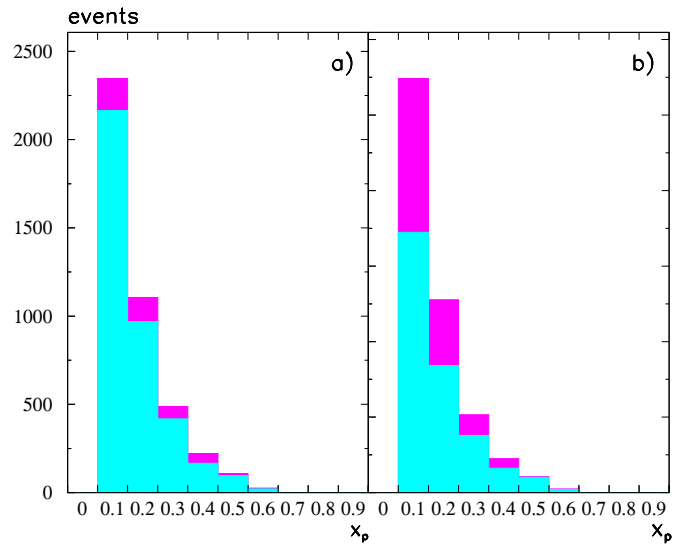}
\caption{The sample of 8000 events for the 
$\bar{p} \, p^\uparrow \rightarrow \mu^+\, \mu^- \,X$ process where an antiproton beam 
with energy $E_{\bar{p}} = 15$ GeV collides on a proton beam with energy $E_p = 3.3$ GeV 
producing muon pairs with invariant mass $4 \leq M \leq 9$ GeV and transverse 
momentum $q_{_T} > 1$ GeV/$c$ (for further details on the cutoffs, see text). The content
of each panel and the notations are as in Fig.~\protect{\ref{fig:hst-40}}.}
\label{fig:hst-coll}
\end{figure}

The lowest range in the muon pair invariant mass offers also a faster rate in the
collection of "good" events. From Tab.~\ref{tab:totxsect}, we deduce that with the above 
mentioned luminosity ${\cal L} = 10^{31}$ cm$^{-2}$s$^{-1}$ it should be possible to
collect a sample 40 times bigger than in the previous case, namely 20000 "good"
events/month; or, equivalently, to reach the same counts (500 "good"  events/month) with 
${\cal L} = 2.5 \times 10^{29}$ cm$^{-2}$s$^{-1}$.

\begin{figure}[t]
\centering
\includegraphics[width=11cm]{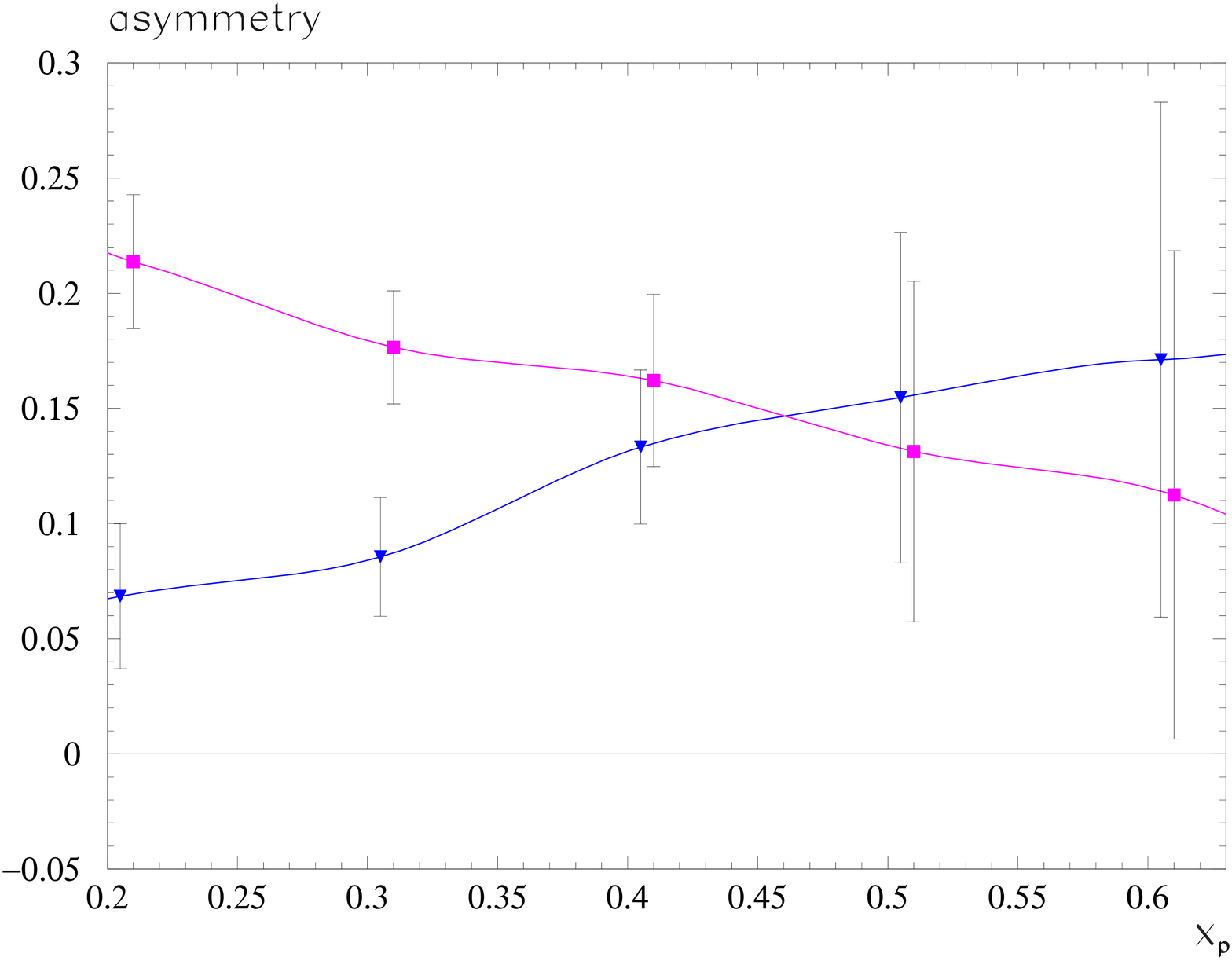}
\caption{Azimuthal asymmetry collected in $x_p$ bins in the same kinematics as in
Fig.~\protect{\ref{fig:hst-coll}} and with the same notations as in
Fig.~\protect{\ref{fig:as-40}}. Continuous lines are drawn to guide the eye. Error bars
from statistical errors only.}
\label{fig:as-coll}
\end{figure}

From previous examples, it seems evident that the smaller the lower cutoff in $\tau$, the
lower the range in $x_p$ where the highest density of events gives the best available 
statistics to extract as many details as possible about $h_1$. Since $\tau = M^2 / s$ and we
want no overlap with $M$ values corresponding to hadronic resonances whose elementary
mechanism for the production of the final muon pair is not well under control, it might be
convenient to considerably increase $s$. The most effective way is to consider the proton
not as a fixed target but as a beam colliding against the antiproton one. Keeping for the
latter the energy $E_{\bar{p}} = 15$ GeV, a proton beam with $E_p = 3.3$ GeV produces an
available cm squared energy $s = 200$ GeV$^2$ (see Sec.~\ref{sec:pT} for details). In
Fig.~\ref{fig:hst-coll}, a sample of 8000 "good" events for the 
$\bar{p} \, p^\uparrow \rightarrow \mu^+\, \mu^- \,X$ process leading to muon pairs with
invariant mass $4 \leq M \leq 9$ GeV is plotted in $x_p$ bins, all the notations and
the other cutoffs being the same as before. Evidently, events are concentrated at lower
$x_p$ bins; but also the asymmetry is, as it can be deduced from the different color codes
of the histograms. In fact, in Fig.~\ref{fig:as-coll} the corresponding azimuthal asymmetry
$(U-D)/(U+D)$ is plotted again with the same notations as in Fig.~\ref{fig:as-40}, the $x_p$
bins being limited to 0.6 because the histograms corresponding to higher $x_p$ values are 
empty. First of all, the 
absolute size of such asymmetry is much bigger (note the different scale with respect to 
Figs.~\ref{fig:as-40} and \ref{fig:as-15}); on the average, it is approximately 15\%. But, 
most important, the best statistics is achieved for $x_p \lesssim 0.4$, where the small 
error bars allow for a very clean distinction between the results corresponding to the 
opposite trends for the function $\langle h_1(x_p) \rangle / \langle f_1(x_p) \rangle$. And 
all this can be obtained with a reduced sample of just 8000 "good" events. 

In conclusion, although the option of a fixed proton target permits a sufficiently clean 
extraction of the analytic properties of the function $h_1(x)$ for $x\lesssim 0.4$ provided
that the invariant mass of the final muon pair is enough low, the collider mode seems more
promising. The higher available cm energy grants a cleaner theoretical description of the
elementary mechanism, gives bigger azimuthal asymmetries, and allows for a much better
statistics and a clean information in the same range $x\lesssim 0.4$. As for the event rate,
from Tab.~\ref{tab:totxsect} we deduce that the luminosity ${\cal L} = 10^{31}$ cm$^{-2}$
s$^{-1}$ would allow to collect 2500 "good" events/month.

\begin{figure}[h]
\centering
\includegraphics[width=10cm]{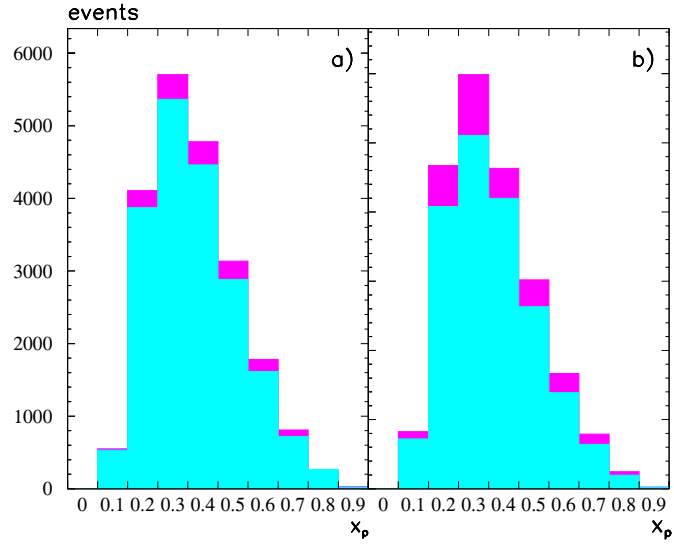}
\caption{The sample of 40000 events for the unpolarized 
$\bar{p} \, p \rightarrow \mu^+\, \mu^- \,X$ process on a fixed $NH_3$ target
with an antiproton beam energy $E_{\bar{p}} = 15$ GeV and with a muon pair of 
invariant mass $1.5 \leq M \leq 2.5$ GeV (for further details on the cutoffs, see 
text). a) Left panel for $1 < p_{_T} < 2$ GeV/$c$. b) Right panel for 
$2 < p_{_T} < 3$ GeV/$c$. For each bin, the darker histogram corresponds to positive values 
of $\cos 2\phi$ in Eq.~(\protect{\ref{eq:mcS}}), the superimposed lighter one to negative 
values.}
\label{fig:hst-unpol}
\end{figure}

\subsection{Unpolarized Drell-Yan}
\label{sec:unpolDY}

In Fig.~\ref{fig:hst-unpol}, the sample of 40000 "good" events for the unpolarized 
$\bar{p} \, p \rightarrow \mu^+\, \mu^- \,X$ process is displayed for the case of a fixed 
$NH_3$ target and for the antiproton beam energy $E_{\bar{p}} = 15$ GeV, i.e. in the same
initial conditions proposed at the HESR at GSI by the PANDA collaboration~\cite{panda}. The
muon pair invariant mass is constrained by $1.5 \leq M \leq 2.5$ GeV and 60 deg. 
$< \theta < 120$ deg. The left panel corresponds to muon pairs with transverse momentum 
$1 < p_{_T} < 2$ GeV/$c$, while the right one to the choice $2 < p_{_T} < 3$ GeV/$c$. The
goal is to test the simple $q_{_T}$ distribution assumed for the coefficient $\nu$ in
Eq.~(\ref{eq:mcnu}) which leads to the corresponding factorized contribution in
Eq.~(\ref{eq:mcS}). For each bin, two groups of events are stored corresponding to positive 
(darker histograms, $U$) and negative (superimposed lighter histograms, $D$) values of 
$\cos 2\phi$ in Eq.~(\ref{eq:mcS}). Again, as in previous plots results are reported in 
$x_p$ bins over the entire range, but the bins at the boundaries, corresponding to 
$x_p < 0.2$ and $x_p > 0.8$, are scarcely or not at all populated and they will be 
discarded in the plot of the corresponding azimuthal asymmetry $(U-D)/(U+D)$.

\begin{figure}[h]
\centering
\includegraphics[width=11cm]{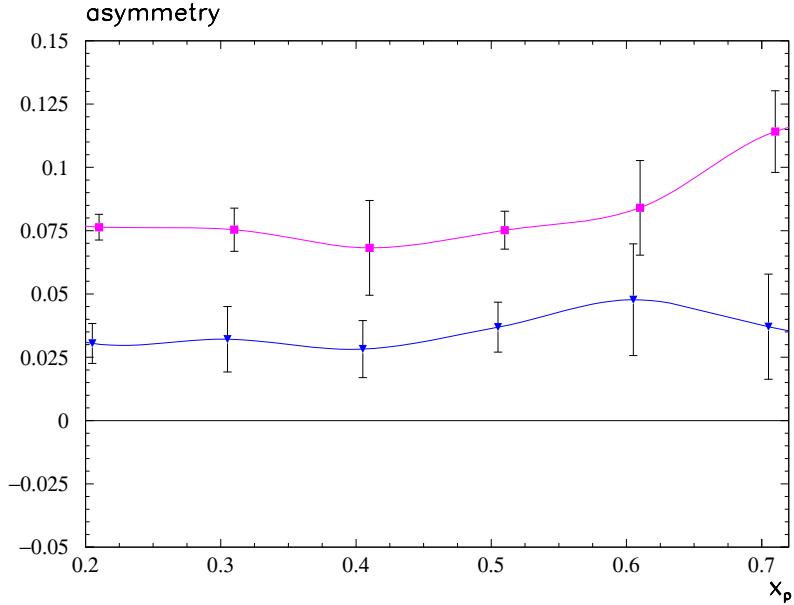}
\caption{Azimuthal asymmetry $(U-D)/(U+D)$ between cross sections in the previous figure
corresponding to the darker histograms ($U$) and superimposed lighter histograms ($D$), 
as bins in $x_p$. Downward triangles for muon pairs with transverse momentum $1<p_{_T}<2$
GeV/$c$, squares for $2<p_{_T}<3$ GeV/$c$. Continuous lines are drawn to guide the eye. 
Error bars due to statistical errors only.}
\label{fig:as-unpol}
\end{figure}

In Fig.~\ref{fig:as-unpol}, the resulting asymmetry $(U-D)/(U+D)$ between cross
sections with positive ($U$) and negative ($D$) values of $\cos 2\phi$ in
Eq.~(\ref{eq:mcS}), is sorted in $x_p$. This asymmetry, already observed at higher 
energy~\cite{Conway:1989fs}, is responsible for the violation of the so-called Lam-Tung sum 
rule and, according to Eq.~(\ref{eq:unpolxsect}), it could be related to the distribution 
function $h_1^\perp$, namely it could be interpreted as a distortion of the parton momentum 
distribution due to its transverse polarization. The predictions of Fig.~\ref{fig:as-unpol} 
should be reliable, because the function $\nu (q_{_T})$ has been tested against the most
recent available measurements~\cite{Conway:1989fs}. In fact, despite its simple functional 
dependence $\nu (q_{_T})$ should not depend on the energy; therefore, even with an 
antiproton beam of 15 GeV the expected size of the asymmetry should be the one displayed in 
Fig.~\ref{fig:as-unpol}. Moreover, in the unpolarized Drell-Yan the magnitude of the 
asymmetry should also not depend on the target, because there is no dilution factor as it 
is the case for the corresponding polarized process. Nevertheless, in
Fig.~\ref{fig:as-unpol} two different sets of asymmetries are plotted corresponding to
different ranges for the transverse momentum of the final muon pair: downward triangles for
$1<p_{_T}<2$ GeV/$c$ and squares for $2<p_{_T}<3$ GeV/$c$. The good statistics and the small
error bars should allow to reliably test the $q_{_T}$ dependence assumed for the coefficient
$\nu$ in Eq.~(\ref{eq:mcnu}).

\section{Conclusions}
\label{sec:end}

In this paper, we have concentrated on the investigation of the spin structure of the 
proton using the single-polarized Drell-Yan process 
$\bar{p} p^\uparrow \rightarrow l^+ l^- X$. At leading twist, the polarized part of the 
cross section contains three contributions with different dependences on the azimuthal 
angle $\phi$ identifying the direction of the final muon pair~\cite{Boer:1999mm}. We have 
focussed on the one involving the convolution of the transversity distribution, the missing 
piece necessary to complete the knowledge of the nucleon spin structure at leading 
twist~\cite{Artru:1990zv,Jaffe:1991kp}, and the distribution $h_1^\perp$, which describes 
how in an unpolarized hadron the momentum distribution of a parton is distorted by its 
transverse polarization. Extraction of the latter is of great importance as well, because 
$h_1^\perp$ contains information on the orbital angular momentum of partons and it appears 
also at leading twist in the unpolarized part of the cross section, weighted by 
$\cos 2\phi$. As such, it is believed to be responsible for the well known 
violation of the Lam-Tung sum rule~\cite{Falciano:1986wk,Guanziroli:1987rp,Conway:1989fs}, 
an anomalous big azimuthal asymmetry of the unpolarized Drell-Yan cross 
section, that neither NLO QCD calculations~\cite{Brandenburg:1993cj}, nor higher 
twists or factorization-breaking terms in NLO 
QCD~\cite{Brandenburg:1994wf,Eskola:1994py,Berger:1979du} are able to justify. Measuring at
the same time the azimuthal asymmetries of the unpolarized and single-polarized Drell-Yan
cross section allows to determine both the unknown functions, i.e. the transversity and
$h_1^\perp$. 

Using unpolarized antiproton beams offers the advantage of involving (anti)parton valence
distributions that should not be suppressed as in the standard fully polarized 
Drell-Yan~\cite{Martin:1998rz,Barone:1997mj}, but avoiding at the same time the difficulty 
of polarizing antiprotons. At GSI a new facility is under development, the High Energy 
Storage Ring (HESR), where an antiproton source will be built. The transversely 
polarized proton target could be obtained from a $NH_3$ molecule where each $H$ nucleus, for
example, is 85\% transversely polarized; a further 25\% dilution factor needs to be applied 
to filter out spurious unpolarized events from the $N$ nucleus. We have simulated
the above single-polarized Drell-Yan process using an antiproton beam of 40 GeV and 
selecting muon pairs with invariant mass $M$ in the range $4 \div 9$ GeV and with a total 
transverse momentum above 1 GeV/$c$. In this kinematics, the quark-antiquark pair 
annihilates into a virtual photon and the theoretical description of the elementary process 
is well under control. We have also made simulations at the lower beam energy of 15 GeV but
with invariant masses in the range $1.5 \div 2.5$ GeV, in order to keep a reasonably large
phase space and at the same time to avoid the overlap with hadronic resonances like the
$\phi$ at 1 GeV and the $J/\psi$ at 3.1 GeV. For such low $M$ values, the parton model
interpretation of the elementary annihilation is questionable because of the contributions
of higher twists. But from the analysis of the scatter plot events seem to concentrate at 
the lowest possible values of $\tau = x_p x_{\bar{p}} = M^2 / s$ (where $x_p, x_{\bar{p}},$ 
are the fractions of hadron momenta carried by the annihilating quark-antiquark pair and 
$s$ is the available initial squared energy in the center-of-mass system); hence, the 
simulation is still useful to study the gross features of the cross section and to explore 
the possibility of extracting the transversity distribution at this lower energy. For both 
beam energies, the polar-angle distribution of the muon pairs is restricted to the range 
$60 \div 120$ deg., where most of the Monte-Carlo events are concentrated. An alternative
way to reach the high-statistics portion of the phase space at low $\tau$ is to increase 
$s$. Therefore, we have considered also the option where the antiproton beam with energy 
$E_{\bar{p}} = 15$ GeV collides on a proton beam with $E_p = 3.3$ GeV such that $s = 200$ 
GeV$^2$. In this last case, $4 \leq M \leq 9$ GeV, all the other applicable cutoffs
being unchanged. 

The primary goal of the simulation is to estimate the number of events required in order to 
access unambiguous information on the distribution functions of interest. Azimuthal
asymmetries have been built, for each bin in the partonic momentum fraction, by separating 
the cross sections with positive values in the $\phi$ dependence from the negative ones and
taking the ratio between the difference and the sum of such values. The $\cos 2\phi$
asymmetry in the unpolarized cross section has already been measured~\cite{Conway:1989fs} at
higher energies. However, under the present kinematical conditions the magnitude of the 
asymmetry should be independent from the energy of the experiment and it can be used as a
cross-check of the Monte-Carlo simulation. It turns out that with an initial sample of 
40000 events (surviving the filter of the various discussed cutoffs) it is possible 
to study in detail the function $\nu$ responsible for such an asymmetry, but in the 
restricted range $0.2 \div 0.8$ for the parton momentum fraction, because the boundary bins 
are not enough populated. 

In the case where antiprotons with energy 40 GeV hit transversely polarized fixed 
proton targets and produce muon pairs with $4 \leq M \leq 9$ GeV, the 40000 
events are enough to produce a statistically nonvanishing azimuthal asymmetry of about 5\% 
on the average, but it seems difficult to extract more detailed information on the
analytic structure of $h_1(x)$ except for very high $x \gtrsim 0.6$, where the adopted QPM 
picture becomes questionable. The lower invariant mass range $1.5 \leq M \leq 2.5$ 
GeV allows to reach lower values of $\tau$ in the phase space, where most of the events are 
concentrated, even for lower antiproton beam energies such as 15 GeV. Hence, a better 
statistics is reached for lower $x_p$ values and it seems possible both to extract the 
functional dependence of $h_1(x)$ for $x \lesssim 0.4$, and to collect events at a higher 
rate (several times bigger than the previous case at the same luminosity). However, the best 
option seems the one where antiproton and proton beams collide reaching high values of $s$. 
In the explored case $s=200$ GeV$^2$, the theoretical description of the elementary 
mechanism is clean and is not affected by complications like, e.g., higher twists; moreover, 
the absolute size of the azimuthal asymmetry turns out bigger and the much better statistics 
should allow for an unambiguous extraction of $h_1(x)$ for $x \lesssim 0.4$ with a reduced 
sample of just 8000 events. 

In conclusion, we hope that the present work will help in the feasibility studies about the
physics program of hadronic collisions with antiproton beams at the HESR at GSI.





\bibliographystyle{apsrev}
\bibliography{mybiblio}

\end{document}